\documentclass[pre,aps,twocolumn]{revtex4}

\usepackage{epsf,graphicx,amsfonts,amsmath,amssymb}
\usepackage{soul}
\usepackage[latin1]{inputenc}
\usepackage[draft]{changes}
\usepackage[draft,textsize=tiny]{todonotes}

\newcommand{\mb}[1]{\mbox{\bfseries \itshape #1}}

\begin{document}

\title{A nonlinear graph-based theory for dynamical network observability}

\author{Christophe Letellier$^1$, Irene Sendi\~na-Nadal$^{2,3}$ \&
Luis A. Aguirre$^4$ } 

\affiliation{$^1$ Normandie Universit\'e CORIA,
Campus Universitaire du Madrillet, F-76800 Saint-Etienne du Rouvray, France}

\affiliation{$^2$ Complex Systems Group \& GISC, Universidad Rey Juan Carlos,
28933 M\'ostoles, Madrid, Spain}

\affiliation{$^3$Center for Biomedical Technology, Universidad Polit\'ecnica
de Madrid, 28223 Pozuelo de Alarc\'on, Madrid, Spain}

\affiliation{$^4$ Departamento de Engenharia Eletr\^onica, 
Universidade Federal de Minas Gerais --
Av. Ant\^onio Carlos 6627, 31.270-901 Belo Horizonte MG, Brazil}

\begin{abstract}
A faithful description of the state of a complex dynamical network would 
require, in principle, the measurement of all its $d$ variables, an unfeasible 
task 
for systems with practical limited access and composed of many nodes with high 
dimensional dynamics. However, even if the network dynamics is observable from 
a reduced set of measured variables,  how to reliably identifying such a 
minimum set of variables providing full observability remains an unsolved 
problem. From the Jacobian matrix of the governing equations of nonlinear 
systems, we construct a {\it pruned fluence graph} in which the nodes are the 
state variables and the links represent {\it only the linear} dynamical 
interdependences encoded in the Jacobian matrix after ignoring nonlinear 
relationships. From this graph, we identify the largest connected sub-graphs 
where there is a path from every node to every other node and there are not 
outcoming links. In each one of those sub-graphs, at least one node must be 
measured to correctly monitor the state of the system in a 
$d$-dimensional reconstructed space. Our procedure is here 
validated by investigating large-dimensional reaction networks for which the 
determinant of the observability matrix can be rigorously 
computed. 
\end{abstract}

\date{
\today}

\maketitle


When dealing with large complex systems, observability
becomes a key concept that addresses the ability of examining the system
dynamics from a reduced set of measurements collected in a finite
time. Indeed, to properly understand the functioning of many
biological or technological networks, it is fundamental to be able to retrieve 
the complex behavior emerging from the local interactions of dynamical
units when just a limited amount of information is available. 

The idea of observability was first introduced by Kalman for linear systems
\cite{Kal60} which was further extended to nonlinear systems by several
other researchers, e.g. \cite{Her77}. That now classical way of
investigating observability provides a yes-or-no answer, that is, the 
system is either fully observable or not through a given set of
measurements. In order to bypass this binary classification of
observability, Friedland proposed the use of a conditioning
number between 0 (non-observable) and 1 (fully observable) to quantify
the observability of linear systems \cite{Fri75}. Later on, Aguirre
showed that observability depends on the chosen coordinate set
to describe the system dynamics \cite{Agu95}. This work led to the
introduction of the observability coefficients to
characterize the observability of  many low-dimensional
chaotic systems \cite{Let98b,Let02,Let05b,Let05a}.

Recently, an attempt to apply those coefficients to small dynamical
networks was reported \cite{Wha15} but, it was pointed out that such
an assesment is out of scope for large dynamical systems due to the
impossibility of calculating the determinant of the corresponding
observability matrix \cite{Bia15}. One way to tackle this drawback is by 
introducing symbolic observability
coefficients \cite{Let09,Bia15} that allow treating larger
dimensional systems although the number of variable combinations to
investigate increases exponentially with the system dimension. 

In order to avoid the use of a brute-force search for a minimum sensor set, 
observability is addressed in \cite{Liu13} by means of graph-theoretic methods. 
Mainly based on performing a linearization of the system (all dynamical 
interdependences between variables are considered constant as in linear 
systems), such a technique reveals that sparse networks with heterogeneous 
degree distributions
are less observable while the observability of denser and homogeneous
networks relies on just a few nodes \cite{Liu13}. However, these latter results
may not hold for nonlinear systems as
the presence of nonlinearities are one of the main causes of
observability loss \cite{Let05b, Bia15}. Some 
variants of this graphical approach were developed in
\cite{Wan14,Lei17} by considering the effect of connection types in
the resulting topologies and in the change of the number of the
necessary sensors. However, none of them actually takes into account
the nonlinear nature of the dynamical interdependence between the
state variables. 
Moreover, it was recently shown that Liu and coworkers' graphical 
approach may not provide the right reduced set of variables to measure (see
the supplement material in Ref.~\cite{Let17} and Ref.~\cite{Hab17}).

Our goal is therefore to address the observability of a complex system 
to identify a minimum set of variables providing access to the rest of
state variables, following an analogous
graph-theoretic approach as in Liu and coworkers (and inspired in
structured system theory \cite{Ros70,Kai80}), but properly handling the
effect of nonlinear dynamical interdependences among variables in the
system's observability. We show the correctness of this approach 
by using benchmark reaction networks coming from biology or
physics and by comparing the obtained results with rigorous algebraic
computations of the determinant of the observability matrix. Our results 
contradict the conclusions drawn in \cite{Liu13} evidencing important
discrepancies mainly resulting from treating linear and nonlinear interdependences
on an equal footing \cite{Let17}.

Let us start by considering a dynamical system whose variables 
$x_i,~i=1,\,2,\ldots, d$ evolve according to 
\begin{equation}
  \label{system}
\dot{ x}_i= f_i({\mb x}),\  
\end{equation}

\noindent
where ${\mb x} \in \mathbb{R}^d$ is the state vector, 
and $f_i$ is the $i$th component of the vector 
field $\mb{f}$. The dynamical system (\ref{system}) is said to be {\it state 
observable} at time $t$ if the initial state $\mb{x} (0)$ can be uniquely 
determined from the knowledge of a variable   $\mb{s} =\mb{h} (\mb{x}) \in 
\mathbb{R}^m$, with $m<d$, measured in the inverval $[0; t]$ \cite{Kai80}. In 
practice, the observability of (\ref{system}) through $\mb{s}$ is assessed 
by computing the rank of the observability matrix 
%
\begin{equation}\label{observabilitymatrix}
  {\cal O}_{\mb s}(\mb x) = 
  \left[
    \begin{array}{l}
      {\rm d} {\mb h}(\mb{x}) \\[0.1cm]
      {\rm d} {\cal L}_{\mb{f}} {\mb h}(\mb{x}) \\[0.1cm]
      \vdots \\[0.1cm]
      {\rm d} {\cal L}^{d-1}_{\mb{f}} {\mb h}(\mb{x}) 
    \end{array}
  \right],
\end{equation}

\noindent
where ${\rm d}\equiv \frac{\partial}{\partial \mb x}$ and 
${\cal L}_{\mb{f}}{\mb h}(\mb x)$ is the Lie derivative of ${\mb h}$ along the 
vector field $\mb f$. This is thus the Jacobian matrix of the
Lie derivatives of $\mb{s}$ \cite{Her77}. The system 
(\ref{system}) is said to be state observable if and only if  the observability 
matrix has full rank, that is, rank$({\cal O}_{ \mb{s}})=d$ \footnote{The observability matrix ${\cal O}_{\mb{ s}}$
  corresponds in fact to the Jacobian matrix of the change of coordinates $\Phi_{{s}} : \mb{x} \rightarrow  \mb{X}$  
where $\mb{X} \in \mathbb{R}^d$ is the reconstructed state vector  from the 
$m$ measured variables and their adequately chosen $d-m$  Lie 
derivatives \cite{Let05a}. }.  Notice that, the full 
observability of a system is determined by the space spanned not only by the 
measured variables  but also by their appropriate Lie derivatives
\cite{Agu05}. 

A systematic check of all the possible combinations
turns out to be a daunting task for large $d$. Therefore, it becomes
crucial to furnish methods to unveil a tractable set of variables
providing full observability of a system. A first attempt was reported
in Liu {\it et al.} \cite{Liu13} using a graphical representation of the functional
relationship among the system variables. We follow such an approach
by choosing as the network representation of the system (\ref{system})
its corresponding ``fluence graph'' where a directed link
$x_j\rightarrow x_i$ is drawn whenever $x_j$ appears in the
differential equation of $x_i$, that is, if the element $J_{ij}$ of the
Jacobian matrix of the Eq.~(\ref{system}) is non-zero \footnote{In \cite{Liu13}
they flip the direction of each edge using the ``inference diagram'' by drawing a directed link
$x_i\rightarrow x_j$ if $x_j$ appears in $x_i$'s differencial equation. }. 

\begin{figure}
  \centering
  \begin{tabular}{ccc}
    \includegraphics[width=0.15\textwidth]{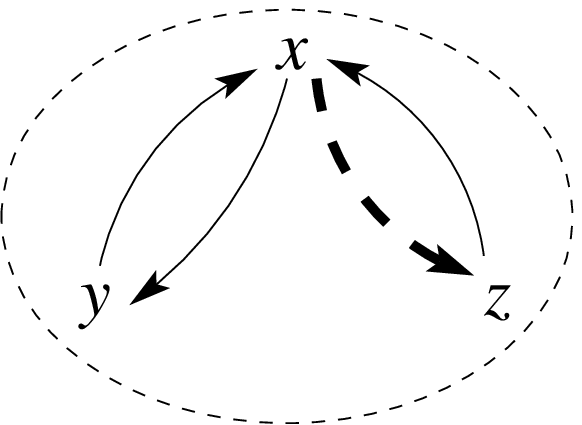} & ~~ &
    \includegraphics[width=0.15\textwidth]{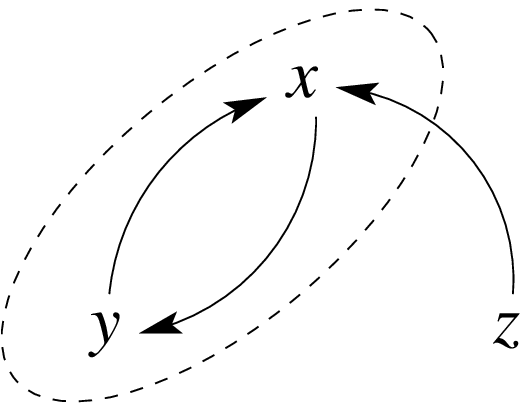} \\[-0.0cm]
    (a) Fluence graph & & (b) Pruned fluence graph \\[-0.2cm]
  \end{tabular}
  \caption{Fluence graphs of the R\"ossler system. (a) Full fluence graph where 
an edge is plotted between variables $x_i$ and $x_j$ whenever $J_{ij}\neq 0$. 
The thick dashed line indicates a nonlinear term in the Jacobian matrix. 
(b) The same as in (a) but edges nonlinearly relating two variables are removed 
from the full fluence graph. A dashed circle surrounds a root strongly 
connected component (SCC). In both graphs, edges $x_i \rightarrow x_i$ are 
omitted since they do not contribute to the determination of the SCC. }
  \label{fig1}
\end{figure}

An illustrative example is provided in Fig.~\ref{fig1}(a) for the R\"ossler 
system ($\mb{x}\equiv (x,y,z)$ and $\mb{f}\equiv (-y-z,x+ay,b+z(x-c))$).  In 
this procedure, a link from $x_j$ to $x_i$ (with $i\neq j$) is present whenever 
$J_{ij}\neq 0$ independently on the linear (solid lines) or nonlinear (thick 
dashed lines) nature of the functional dependence.  At this point is where we 
deviate from Liu and coworkers approach as it ignores the fact that a lack of 
observability most often originates from the nonlinear 
relationship between variables \cite{Let05b}. In order to correct this 
shortcoming, we propose to distinguish linear from nonlinear couplings
\cite{Let09,Bia15} by {\it pruning from the fluence graph all nonlinear
links} and keeping only those associated with the constant elements in the 
Jacobian matrix of the system. We call this reduced fluence graph, the {\it pruned fluence graph} 
(Fig.~\ref{fig1}(b)) that we take as the minimum graph containing the 
information flow that will allow us to select the minimum set of sensors to 
ensure observability of the whole system while working in a 
$d$-dimensional reconstructed space.

\begin{figure}[ht]
  \begin{tabular}{cc}
    \multicolumn{2}{c}{
      \includegraphics[width=0.15\textwidth]{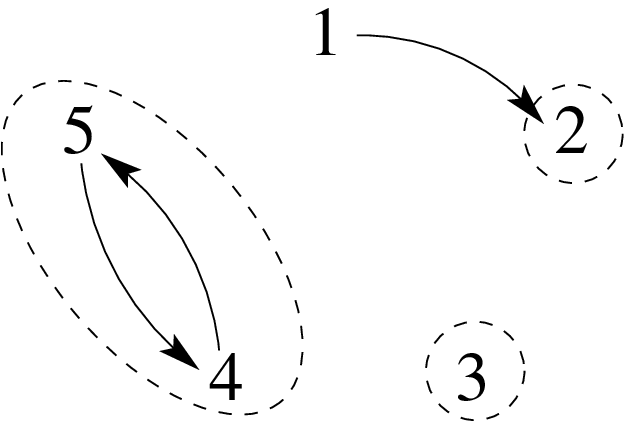}} \\   
    \multicolumn{2}{c}{(a) 5D rational system} \\[0.2cm]
    \includegraphics[width=0.22\textwidth]{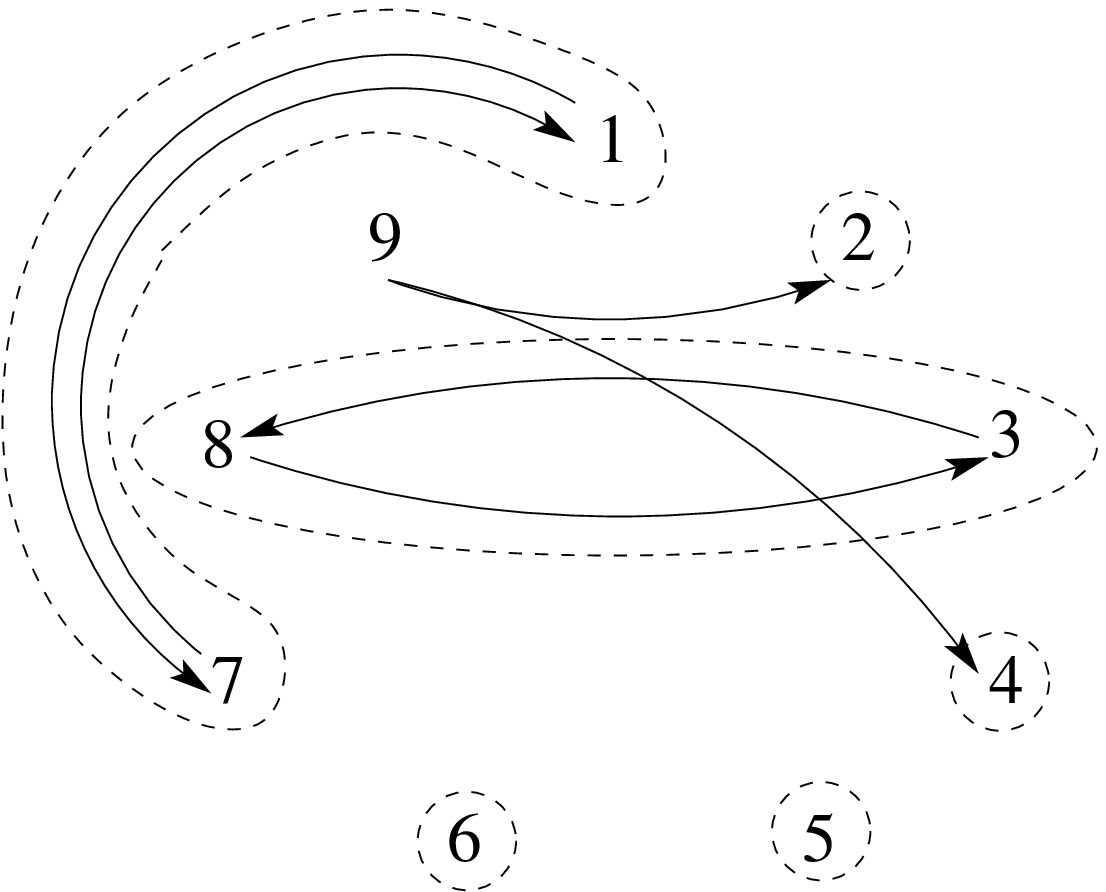} &
    \includegraphics[width=0.22\textwidth]{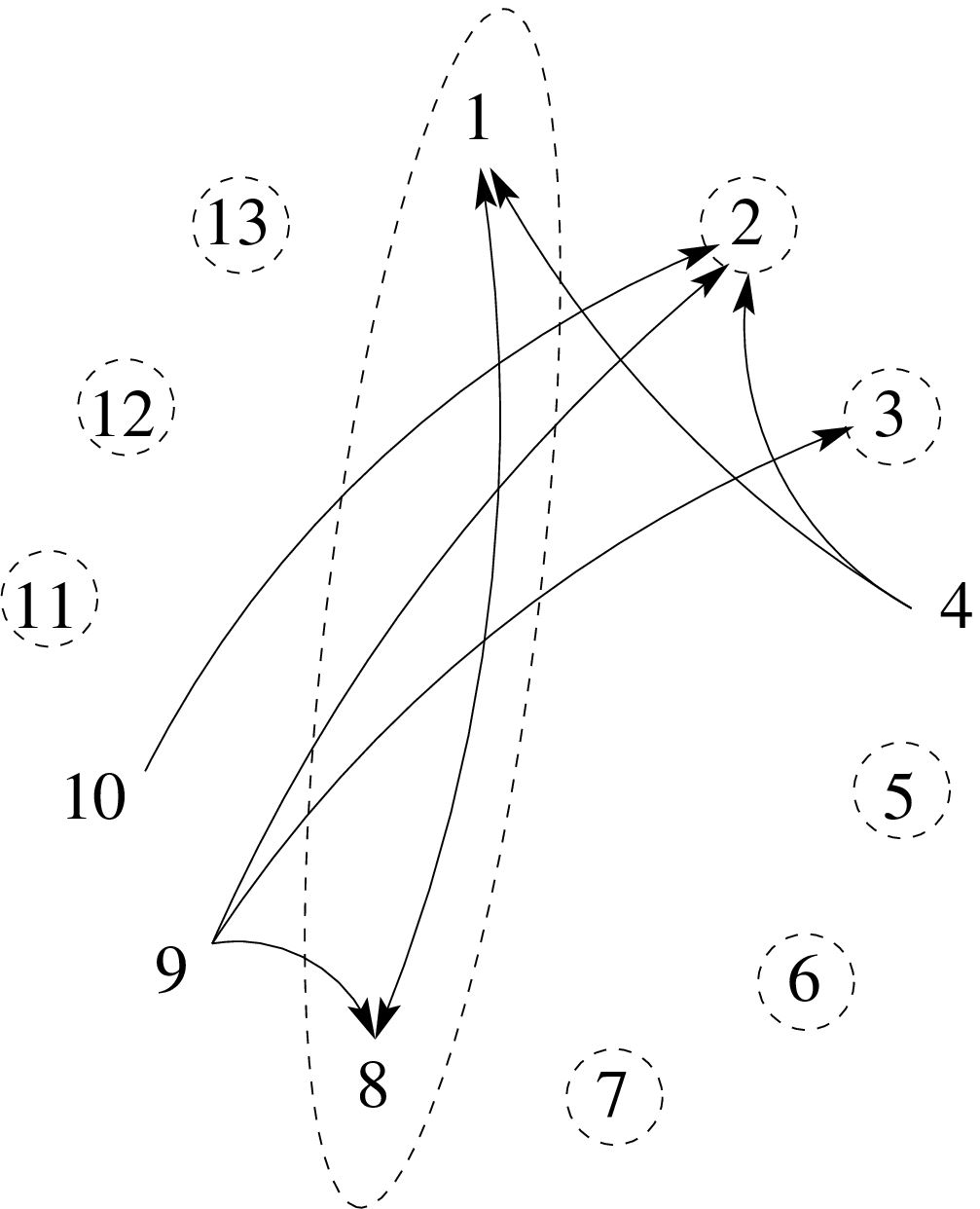}  \\
    (b) 9D RB system & (c) 13D DNA system \\[-0.2cm]
  \end{tabular}
  \caption{Pruned fluence graphs of (a) the 5D rational system for the 
circadian oscillations in the Drosophila period protein, (b) the 9D 
Rayleigh-B\'enard convection, and (c) the 13D reaction network for the  
replication of fission yeast. Numbers $i$ label the variable $x_i$ of the 
models, continuous lines from $i$ to $j$ represent that variable $x_j$ is 
linearly influencing variable $x_i$ and variables surrounded by a dashed circle 
are part of an SCC without outcoming edges.  
  \label{fig2}}
\end{figure}

In the following, the graph analysis described in \cite{Liu13} to
isolate the minimum set of sensors still holds. Namely, a node in the
pruned fluence graph is a sensor if it belongs to a root strongly connected
component (SCC) of the graph (a subgraph in which there is a directed path
from each node to every node in the subgraph) and with no {\it outcoming}
links, that is, such an SCC is either an isolated subgraph or a root (a sink) of 
information flowing from any other subgraphs in the network \footnote{Notice 
that, since in \cite{Liu13} they flip the edge direction of the edges with 
respect to our choice for the graph representation of the dynamical system, 
their definition of root SCC is reversed here.}. By
measuring at least one of the nodes in each subgraph classified as a
root SCC in the pruned fluence graph, we make the conjecture that such
a selection is not only minimal but also provides a good observability. 

In order to validate our hypothesis, we applied the above procedure to several 
nonlinear dynamical systems widely known in the physical and biological 
scientific community. For each of them we confirm that a candidate set of 
variables to be measured actually provides full observability by checking that 
the determinant Det~${\cal O}_{\mb s}$ of the analytical observability matrix 
${\cal O}_{\mb s}$ as defined in Eq.~(\ref{observabilitymatrix}), 
is always nonzero \cite{Let02,Fru12}. Note that for dimensions larger than 4, 
the determinant cannot always be computed due to its complexity (Maple software 
fails to compute some observability matrices for a 5D rational system 
\cite{Bia15}). To deal with this difficulty, a symbolic formalism was 
introduced in \cite{Let09,Bia15} that allows to quantify the observability of a 
given measure by means of a symbolic observability coefficient $\eta=1$, if the 
observability is full, $\eta>0.75$ if good, and poor otherwise \cite{Sen16}.  
Briefly, it is based on a symbolic Jacobian matrix $\tilde J_{ij}$  whose 
elements can be either $1,\bar{1}$ and $\Bar{\Bar{1}}$ which encode, 
respectively, constant, nonlinear and rational terms (whose denominators 
contain variables $x_j$) of the Jacobian matrix 
$J_{ij}=\partial f_i/\partial x_j$ \cite{Bia15}. 

It turns out that the symbolic observability coefficients are inversely 
proportional to the complexity of the determinant of the observability matrix 
\cite{Let02}. Therefore, in general, in those cases in which the sensor set is 
providing full observability, the determinant Det~${\cal O}_{\mb s}$ can thus 
be analytically computed. We used this property as a validation of our 
hypothesis, and check whether a non vanishing determinant is obtained for a 
given set of variables potentially providing full observability. 


\noindent {\it The R\"ossler system.} Let us now consider the R\"ossler system whose Jacobian matrix 
\begin{equation}
  J_{\rm Ros} = 
  \left[
    \begin{array}{ccc}
      0 & -1 & -1 \\
      1 & a & 0 \\
      z & 0 & x-c
    \end{array}
  \right]
\end{equation}
has two non-constant terms and whose symbolic form accounting only for
the linear dynamical interdependencies is reduced to
\begin{equation}\label{SymJacRoss}
  \tilde{J}^{\rm lin}_{\rm Ros} =
  \left[
    \begin{array}{ccc}
      0 & 1 & 1 \\
      1 & 1 & 0 \\
      0 & 0 & 0
    \end{array}
  \right] \, . 
\end{equation}

\noindent
The two nonlinear terms, $J_{31}$ and $J_{33}$ are, therefore, not considered 
for constructing the pruned fluence graph shown in Fig.~\ref{fig1}(b). On the 
other hand, when all terms in the Jacobian matrix are considered equivalent 
independently of their linear or nonlinear nature as in \cite{Liu13}, the 
resulting graph is the one shown in Fig.~\ref{fig1}(a) except that the thick 
dashed
line will be drawn as a thin line. For this latter case, the decomposition in 
SCC singles out a root SCC composed by the three variables, suggesting that any 
of the three variables can be measured to achieve observability. 
However, it is known that measuring variable $z$ alone provides poor 
observability ($\eta_{z,\dot{z},\ddot{z}} = 0.44$) of the R\"ossler dynamics 
\cite{Let98b,Bia15}. The picture changes completely if only the linear 
dependencies are considered as depicted in the pruned fluence graph 
(Fig.~\ref{fig1}(b)). In this case, this graph contains a single root SCC 
composed only of variables $x$ and $y$ for which $\eta_{x,\dot{x},\ddot{x}} = 
0.84$ and $\eta_{y,\dot{y},\ddot{y}} = 1$, respectively. Variable $z$ is 
therefore no longer part of the recommended set of measurements when taking 
into account the pruned fluence graph, in full agreement with the symbolic 
observability coefficients \cite{Let17}.

Moreover, and not surprisingly confirming these results, when looking at the  
determinants of the corresponding analytical observability matrices we find 
that Det~${\cal O}_{y \dot{y} \ddot{y}} = 1$, whose constant value means full
observability, Det~${\cal O}_{x \dot{x} \ddot{x}} = x - (a+c)$, meaning that 
observability is good as long as the determinant is not vanishing and it 
depends on variable $x$ (order 1), and  Det~${\cal O}_{z \dot{z} \ddot{z}} 
= z^2$, indicating that observability is very poor as it depends on the square 
of $z$ (order 2).  

\noindent{\it A 5D model for the circadian oscillations of the Drosophila.} Let 
us now consider a 5D rational model for the circadian oscillations in the 
{\it Drosophila} period protein \cite{Gol95}. Dynamical equations and the 
corresponding symbolic Jacobian matrix reduced to just the constant elements 
are reported in the Supplemental Material for all the systems considered in 
the subsequent part of this letter. The pruned fluence graph (Fig.\ 
\ref{fig2}a) presents three root SCC, thus suggesting that just measuring 
variables $x_2$ and $x_3$ and either $x_4$ or $x_5$, is sufficient to fully and efficiently
account for the dynamics of the whole system. This prediction about the 
appropriate  set of variables to monitor is confirmed by the symbolic 
observability coefficients of the reconstructed spaces $\{x_2^2 x_3 x_5^2\}$ 
and $\{x_2^2 x_3 x_4^2\}$ \footnote{The notation $\{x_i^j\}$ is equivalent to 
the vector $\{x_i,\dot{x_i},\ddot{x_i},...\}$ up to the $j$th time derivative.} 
whose values are both equal to 1, and by the constant analytical determinants 
Det~${\cal O}_{x_2^2 x_3 x_5^2} = -k_s k_2$ and 
Det~${\cal O}_{x_2^2 x_3 x_4^2} = k_s k_1$, where $k_1$, $k_2,$, and $k_s$ are 
parameters of the model \cite{Let17}.

\begin{table*}
  \centering
  \caption{Minimum set of variables that are needed to measure according to 
{\it i)} Liu and coworkers' inference graph, and {\it ii)} the proposed pruned 
fluence graph. The cardinality of both minimum sensor sets $m_1$ and $m_2$ is 
reported in the column next to each case. Last column indicates the required 
number of variables to measure for getting full observability according to 
exact analytical calculations \cite{Let17}.  $\lor \equiv$~or and 
$\land \equiv$~and.}
  \label{compar}
  \begin{tabular}{lccccc}
    \\[-0.3cm]
    \hline \hline
    \\[-0.4cm]
    Model & Inference graph & $m_1$ & Pruned fluence graph &$m_2$   
    & $m \,(\eta =1)$ \\[0.1cm]
    \hline 
    \\[-0.4cm]
     3D & $x \lor y \lor z$ & 1 & $x \lor y$ & 1 & 1 \\[0.1cm]
     \hline
     5D & $x_1 \lor x_2 \lor x_3 \lor x_4 \lor x_5$ & 1  &
     $x_2 \land x_3 \land (x_4 \lor x_5)$ & 3 & 3 \\[0.1cm]    
     \hline
     9D & $x_6 \land (x_1 \lor x_2 \lor x_3 \lor x_4 \lor x_5 
      \lor x_7 \lor x_8 \lor x_9)$ & 2 &  $(x_1 \lor x_7) \land x_2 \land x_4 
     \land x_5 \land x_6 \land (x_3 \lor x_8)$ 
     & 6  & 6 \\[0.1cm]    \hline
     13D  & 
      $(x_1 \lor x_2 \lor x_3 \lor x_4 \lor x_5 \lor x_8 \lor x_9 \lor x_{10})$      & 5 & $(x_1 \lor x_8) \land x_2 \land x_3 \land x_5 \land x_6 \land  x_7 
     \land x_{11} \land x_{12} \land x_{13}$ & 9 & 10 \\[-0.1cm]
     & $ \land x_6 \land x_7 \land x_{11} \land x_{12}$ & &  & & \\[0.1cm]
    \hline \hline
  \end{tabular}
\end{table*}

\noindent{\it A 9D Rayleigh-B\'enard convection model. } Let us now consider a 
9D system describing the Rayleigh-B\'enard convection in a square platform 
\cite{Rei98}. 
Its pruned fluence graph shown in Fig.~\ref{fig2}b exhibits six root SCC, 
suggesting that a good observability might be obtained by measuring variables 
$x_2$, $x_4$, $x_5$, and $x_6$, and either  one between $x_3$ or $x_8$, and 
another between $x_1$ or $x_7$. Therefore, at least 6 variables are needed in 
this 9D reaction network to effectively ensure full observability. This selection of 
variables is in agreement with the fact that most of the combinations whose 
observability coefficient is equal to 1, do not involve variables $x_1$, $x_3$, 
$x_7$, $x_8$, and $x_9$
\cite{Let17}. For instance, 
Det~${\cal O}_{x_1^2 x_2^2 x_3^2 x_4 x_5 x_6} = - b_2^2 \sigma^3/2$ (where 
$b_2$ and $\sigma$ are parameters of the model,
confirming that a full observability can be indeed obtained with this reduced 
set of measured variables. 

\noindent{\it A 13D model for the DNA replication in fission yeast.} A more 
challenging dynamical system, that is also analysed in \cite{Liu13}, is the 
model for cell cycle control in fission yeast governing 
the concentrations of the state variables \cite{Nov97}. 
The corresponding pruned fluence graph (Fig.\ \ref{fig2}c) presents 9 root 
SCCs, meaning that, at least 9 variables --- one from each SCC --- must be 
measured. The detection of the SCCs of the pruned fluence graph tells us 
immediately that variables $x_4$, $x_9$ and $x_{10}$ can be discarded from the 
minimum sensor set as well as that either variable $x_1$ or $x_8$, can be 
excluded but not simultaneously. Indeed, the combinations providing good 
observability ($\eta = 0.93$) with 9 variables measured do not involve the sets
$\{ x_4, x_8, x_9, x_{10} \}$ or  $\{ x_1, x_4, x_9, x_{10} \}$. For instance, 
the reconstructed space spanned by 
$\{x_1^2 x_2^2 x_3^2 x_5 x_6 x_7 x_{11} x_{12}^2 x_{13}\}$ yields a 
$\eta=0.93$ and Det~${\cal O} = \frac{(k_{7r} + k_2 + k_{2'}) (k_{8r} 
+ k_4) (x_{12}-1) (k_{7r} + k_4)}{K_{\rm mc} + 1 - x_{12}} k_{\rm c} \beta$, 
that is, constant over the whole state space but the plane $x_{12} = 1$. This 
indicates that the developed framework to determine the 
minimum sensor set is not guaranteeing full observability in a 
$d$-dimensional reconstructed space but its observation is mandatory to ensure 
good observability. In fact, as reported in \cite{Let17}, an extra variable 
must be added for getting a full observability. For instance, 
$\eta_{x_1^2 x_2 x_3^2 x_5 x_6 x_7 x_8^2 x_{11} x_{12} x_{13}} = 1$ is
associated with Det~${\cal O}_{\mb{s}} = (k_{7r} + k_4)^2 (k_{8r} + k_4)$ which 
never vanishes. These results are in full disagreement with those reported  in 
\cite{Liu13} where the authors build a fluence graph treating all dynamical 
interdependences as linear. Just for illustration, their analysis gives rise to 
the existence of two SCCs: $\{x_{13} \}$ and another one with the rest of the 
12 variables which is a root SCC. Therefore, their conclusion is that by just 
monitoring any variable in the root SCC, the system is observable as verified 
by the Sedoglavic's algorithm \cite{Sed02}, an algorithm that certifies the 
local (not global) observation in a probabilistic way. 

A more thorough comparison of the two approaches is given in 
Table ~\ref{compar}, where the minimum sensor set is reported in each case for 
the four nonlinear dynamical systems considered in this Letter. As observed in 
the last column of the table, where the minimum number $m$ of variables needed 
to get full observability according to exact analytical calculations, Liu and 
coworker's approach tends to underestimate the number of variables.

Complex networks are large dimensional systems for which it is not possible to
measure all the variables required for a full description of any of their
states. 

Considering that nonlinear links are generally responsible for the lack of
local observability, we proposed 
constructing a pruned fluence graph considering only linear links -- corresponding to the constant non zero 
terms of the Jacobian matrix of the network. We showed that identifying 
the root SCCs of the pruned fluence graph allowed to correctly identify the 
reduced set of measurements providing a good observability of the network 
dynamics. This technique was validated with the use of symbolic observability 
coefficients and the analytical determinants of observability matrices. 
We thus presented an easy-to-implement technique for selecting the variables to 
be measured for reconstructing a $d$-dimensional space of a reaction network.
  The extension to networks of  dynamical systems is straightforward as long as the Jacobian matrix describes the node dynamics and
  their connectivity.

\begin{acknowledgments}
ISN acknowledges
partial support  from the Ministerio de Econom\'ia y Competitividad
of Spain under proj ect FIS2013-41057-P and from the Group of Research 
Excelence URJC-Banco de Santander. LAA acknowledges CNPq.
\end{acknowledgments}

\bibliography{references}

\end{document}